\documentclass[aps,pre,twocolumn,epsfig,showpacs]{revtex4-1}
\usepackage{graphics}
\usepackage{subcaption}
\usepackage{graphicx}
\usepackage{epsfig}
\usepackage{amsmath}
\usepackage{epstopdf}
\usepackage{color}
\usepackage{float}

\usepackage{csquotes}
\usepackage{mathtools}

\usepackage{hyphenat}
\usepackage{epstopdf}
\usepackage{times}
\let\oldref\ref
\renewcommand{\ref}[1]{(\oldref{#1})}
\usepackage[colorlinks=true,linkcolor=blue,urlcolor=blue,citecolor=blue]{hyperref}
\begin{document}

\title{Polymer translocation: Effects of confinement}
\author{Manish Dwivedi${^1}$, Sumitra Rudra${^1}$ and Sanjay Kumar${^1}$}
\affiliation{${^1}$Department of Physics, Banaras Hindu University, Varanasi 221005, India }

\begin{abstract}
 We investigate the influence of varying confinement on the  dynamics of polymer translocation through a cone shaped channel. For this,  a linear polymer chain is modeled by a self avoiding walks (SAWs) on a square lattice. The {\it cis}-side of a cone-shaped channel has a finite volume, whereas the {\it trans-}side has a semi-infinite  space. The confining environment is varied  either by changing the position of the back wall while keeping the apex angle fixed or altering the apex angle while keeping the position of the back wall fixed. In both cases, the effective space $\phi$, which represents the number of monomers in a chain relative to the total number of accessible sites  within the cone, is reduced due to the imposed confinement. Consequently, the translocation dynamics are affected. We  analyze  the  entropy of the confined system  as a function $\phi$, which exhibits  non-monotonic behaviour. We also calculate the  free energy associated with the confinement as a function of  a virtual coordinate for different positions of the back wall (base of the cone) along the conical axis for various apex angles. Employing the Fokker-Planck equation, we calculate the translocation time  as a function of $\phi$ for different  solvent conditions across the pore.  Our findings indicate that the translocation time decreases as $\phi$ increases,  but it eventually reaches a saturation point at a certain value of $\phi$. Moreover, we highlight the possibility to control the translocation dynamics by manipulating the solvent quality across the pore. Furthermore, our investigation delves into the intricacies of polymer translocation through a cone-shaped channel, considering both repulsive and neutral interactions with the pore wall. This exploration unveils nuanced dynamics and sheds light on the factors that significantly impact translocation within confined channels.

\end{abstract}

    \maketitle

\section{Introduction}
The translocation of a bio-polymers through a pore is a fundamental step in many cellular processes {\it e.g.}  passage of DNA and RNA through nuclear pores\cite{Vella1994}, protein transport through membrane channels \cite{Nakielny1999}, packaging of genome in bacteriophages, ejection of  virus from capsid etc\cite{Molineux2013,Marenduzzo2013,Berndsen2014}. Apart from it's biological relevance, interest has also been fueled to its biotechnological applications  such as gene therapy, drug delivery, DNA sequencing etc \cite{Kasianowicz1996,Schneider2010}. In fact, recent studies revealed that polymer translocation through pores is a complex process, and is governed by the combined effects of various factors, such as the nature of the pore-polymer interaction \cite{Luo2007,Chen2009,Luo2012}, the geometry of the pore \cite{Polson2019,Mohan2010}, external driving forces\cite{Luo2006}, salt concentration \cite{Dabhade2022,Hsiao2020} across the pore etc. A detailed understanding of the effect of various system parameters on the translocation process, is prerequisite to understand the transport dynamics of the translocating molecule. 

For the unforced translocation,  the mean passage time $\tau$ scales as $N^\alpha$ with $\alpha= 1+2\nu$, where $\nu$ is the Flory exponent. However, recent studies suggest that the value of scaling exponent $\alpha$ depends on the pore size,  apex angle, quality of solvents etc. \cite{Luo2008,Edmonds2012,Ikonen2012,Bhattacharya2009}.
Significant amount of work  have been carried out now to understand the role of driving force on the translocation. In case of driven translocation, there exist several possibilities:  the driving force could be in the form of a chaperon binding \cite{Abdolvahab2011,Suhonen2016,Yu2011}, confinement \cite{Muthukumar2001,Cacciuto2006,Wong2008},  crowders \cite{Polson2019,Khalilian2021,Tan2021} or an external field\cite{Luo2006} . In case of charged polymers, e.g., DNA, the translocation is driven by an electric field acting across the channel. Kasianowicz {\it et al.} \cite{Kasianowicz1996} studied the translocation of single molecules by measurement of the ionic current passing through pores. The potential difference across the pore drives the ionic current, which  also drives charged biopolymers through the pore. The amount of the variation in the ionic current has been used to extract the information about the  sequence and length of DNA. 

\begin{figure}[t]
\centering\includegraphics[scale=0.30]{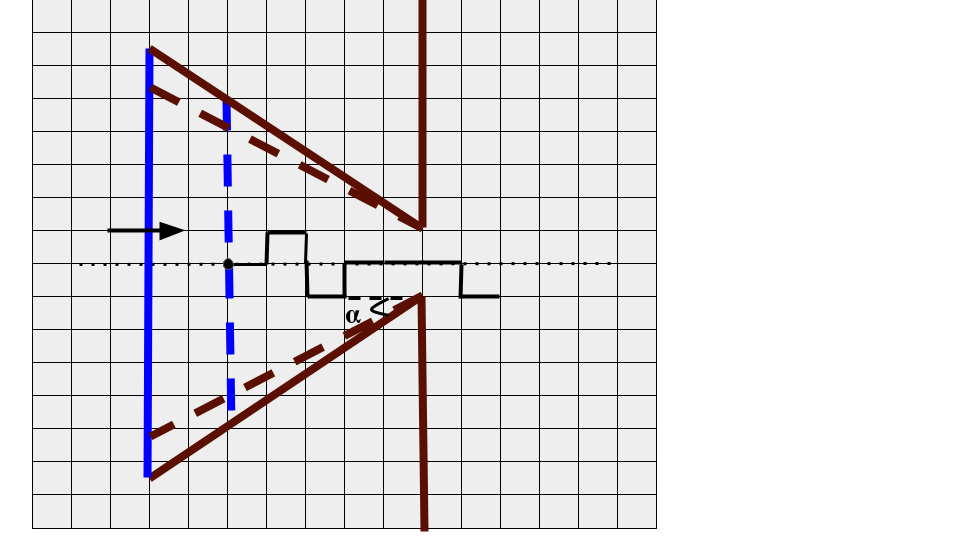}
\caption{ Schematic representation of of a polymer threading through  a conical channel (pore) on a square lattice.  The impenetrable rigid side walls are depicted as thick brown walls, while the impenetrable rigid back wall is shown as a thick blue line. The conical channel separates the cis and trans sides, with infinitely extended rigid walls. The confinement within the pore can be adjusted by either changing the position of the back wall, indicated by the dashed blue line, or by varying the angle of the cone, represented by the dashed brown line. The polymer within the pore undergoes a random walk, starting from various positions on the principal axis, denoted by the dotted line.}
\vspace{-0.1in}
\label{fig1}
\end{figure}

Confinement arising due to the geometry of the pore-interface (e.g. Mycobacterium smegmatis porin A (MspA), HIV-1 capsid) also drives the translocation \cite{Jeon2016}. The geometry of the pore looks similar to a cone-shaped channel Fig.\ref{fig1}, where driving force may be due to the different solvents across the pore or entropy gradient because of the shape of the channel \cite{Nikoofard2012,Sharma2022}.  The polymer translocation through the cone-shaped channel has been studied experimentally \cite{Lan2011, Sexton2007, Siwy2002} and theoretically in recent years. The focus was to obtain scaling associated with translocation \cite{Sung1996, Muthukumar1999, Muthukumar2003, Wong2008, Muthukumar2001}. On the other hand, theoretical understanding of equilibrium properties of a polymer  inside the cone has been achieved through the conformal invariance technique  and the exact enumeration technique followed by the series analysis \cite{Guttmann1984} to calculate various exponents. Using exact enumeration technique followed by the Langevin Dynamics simulations, Kumar {\it et al}\cite{Kumar2020} showed that the confinement leads to the non-monotonic behaviour in the $\theta-$temperature, where a polymer chain acquires the globule state from the coil state. Interestingly, by varying the apex angle $\alpha$, they showed that $\theta-$ temperature can be shifted. 

In this study, we embark on a comprehensive exploration of polymer translocation dynamics with a distinct focus on the influence of the confining environment. What sets our research apart are three pivotal elements: firstly, our model embraces a realistic depiction by featuring a finite volume on the cis side and semi-infinite space on the trans side, closely mirroring real-world systems. Secondly, our model allows for dynamic adjustments in the apex angle and the position of the back wall, potentially inducing shifts in the polymer's center of mass due to changes in geometry. Thirdly, we delve into the entropy variation concerning the back wall position and the apex angle and unveil the free energy profile of the system, offering insights into the thermodynamics of polymer translocation. Finally, we employ the Fokker-Planck equation to calculate the translocation time as a function of the back wall position and the effective space available to the polymer. These novel aspects of our research not only deepen our understanding of polymer translocation but also illuminate the hitherto uncharted territory of confining environment effects on this dynamic process, highlighting the unexplored areas that demand attention within this field. The outline of the paper is as follows:

In Sec.\ref{l1} , we provide a brief overview of the model and elucidate the employed techniques.  Sec. \ref{l2} delves into the calculation of the system's entropy as a function of $\phi$ and the free energy.  Since,  the exact information of density of states is known, we employ the Fokker Planck equation to calculate the time required to cross the pore and termed it as a translocation time ($\tau$). In Sec.\ref{l3}, we   explore the impact of both repulsive and neutral interactions with the pore wall, revealing nuanced dynamics that significantly influence translocation within confined channels. The paper ends with a brief discussion and future perspective of the present study in Sec.\ref{l4}. 
 
\section{Model and method}
\label{l1}
We investigate a self avoiding walk model of a linear polymer chain on a square lattice, which is confined in a cone shaped channel. The cone shaped channel composed of two perfectly reflecting hard walls that are inclined at an apex angle $\alpha$ (Fig. 1).  A single monomer is only allowed to pass through a pore, which prohibits the formation of a fork or hairpin at pore. To explore the process of translocation resulting from reduced entropy due to the closed cone shaped channel,  we employ an exact enumeration technique to systematically generate all possible conformations of the polymer, and examine whether the polymer resides in the {\it cis-}side (within a channel) or {\it trans-}side (out side the channel) or in a state of threading (from the {\it cis-}side to the {\it trans-}side).  The inherent fourfold symmetry of the square lattice naturally produces angles that are multiples of $\pi/4$, such as $\pi/4$ and $\pi/2$. To explore angles beyond these specific multiples, we employed an innovative technique outlined in Ref \cite{Chauhan2021}. Specifically, we consider $n_x$ bonds along the $\pm x$ direction, followed by $+n_y$ bonds in the $y$ direction. Distinguishing our study from previous study \cite{Chauhan2021}, where the {\it cis }and {\it trans}- sides were characterized by semi-infinite space. The model developed here represents a significant departure from previous approaches, reflecting a fundamental shift towards a more realistic representation of real-world systems. The presence of a perfectly reflecting hard back wall in the conical channel (Fig. \ref{fig1}) alters the translocation dynamics. This marked departure from the conventional semi-infinite space assumption may impact our understanding of polymer translocation within confined environments. The position of the back wall can be placed at different positions of the principal axis  (say $X =-18,-17, ....$)  away from the origin. Moreover, the {\it trans-}side is demarcated by a reflecting vertical (perpendicular to the principal axis) hard wall  with a channel (Fig. \ref{fig1}). It's worth emphasizing that alterations to the apex angle or the back wall's position lead to changes in the available space within the {\it cis}-side. However, in stark contrast, the available space within the {\it trans}-side remains constant and unaffected by variations in the confining environment. 

We have attached one end of a SAW at the middle of the back wall placed at a distance $-X$ away from the pore and generated all possible conformations of $N = 28$ steps walk with the imposed boundary conditions. We systematically varied the position of the back wall ($X=-18$ to  $X=-1$), and evaluated the partition function as a simply sum over all possible conformations for different apex angles.  
 \begin{equation}
     Z_X = C_N z^{N},
 \label{1}    
 \end{equation}
 where $\it C_N$ is the total number of all possible conformations of a polymer chain whose one end is attached on the back wall (Fig. \ref{fig1}) and $z$ is the fugacity of each step of the walk. For the large $\it N$, $ \it C_N \sim \mu ^{N} N^{\gamma^ {\prime} -1}$ \cite{Vanderzande1998,Kumar2010}. Here $\mu$ is the connectivity constant of the lattice and $\gamma^{\prime}$ is the critical exponent associated with the SAW confined in a cone of the apex angle $\alpha$. It's important to note that the techniques previously developed and successfully applied in Ref. \cite {Chauhan2021} have established their efficacy in reproducing earlier findings. Building upon this well-established foundation, we have extended these techniques to a more realistic model. This extension allows us to elucidate the fundamental physics governing translocation in scenarios characterized by diverse geometric considerations and confinement factors.
 
\begin{figure}[t]
\includegraphics[scale=0.30]{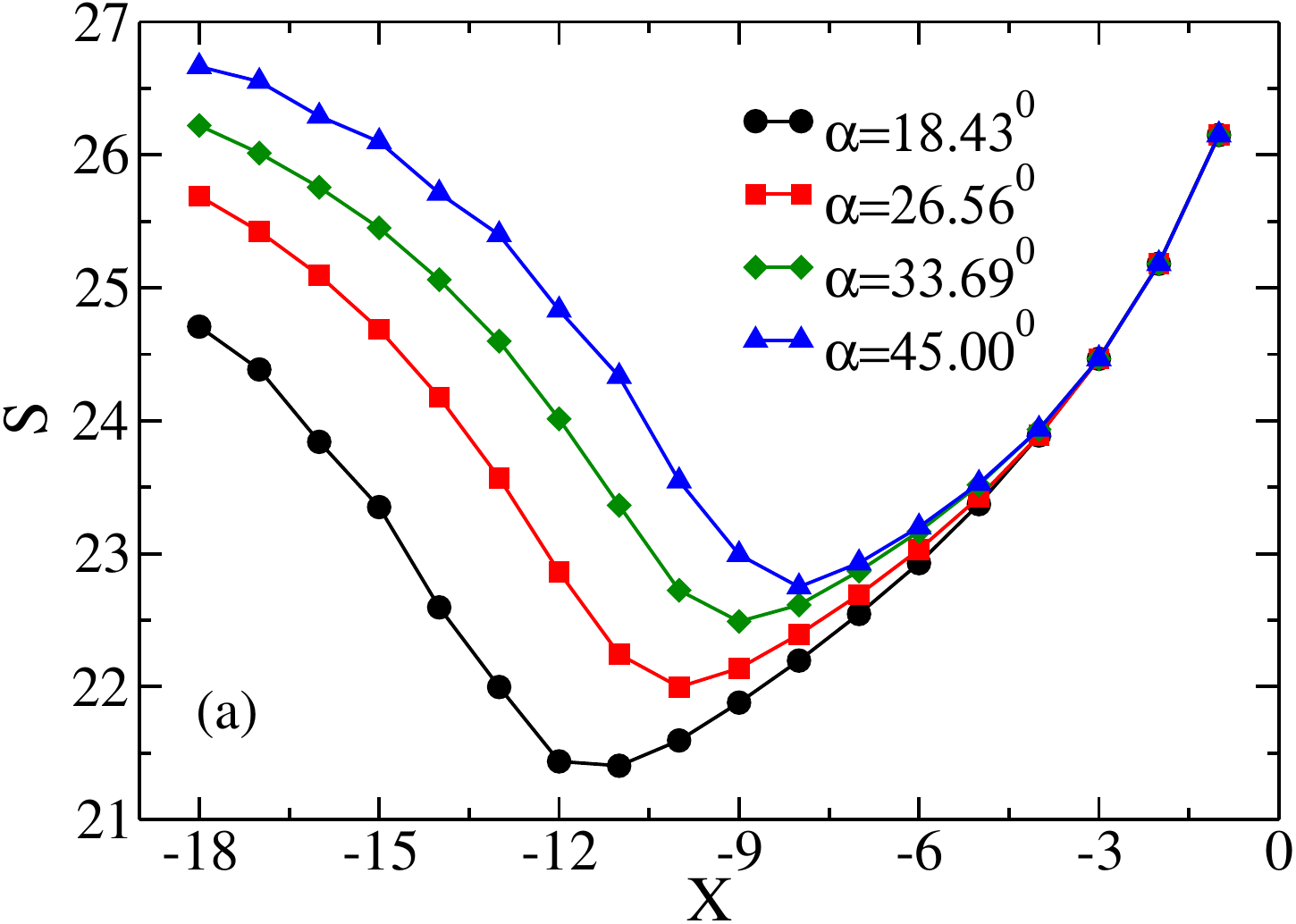}
\includegraphics[scale=0.30]{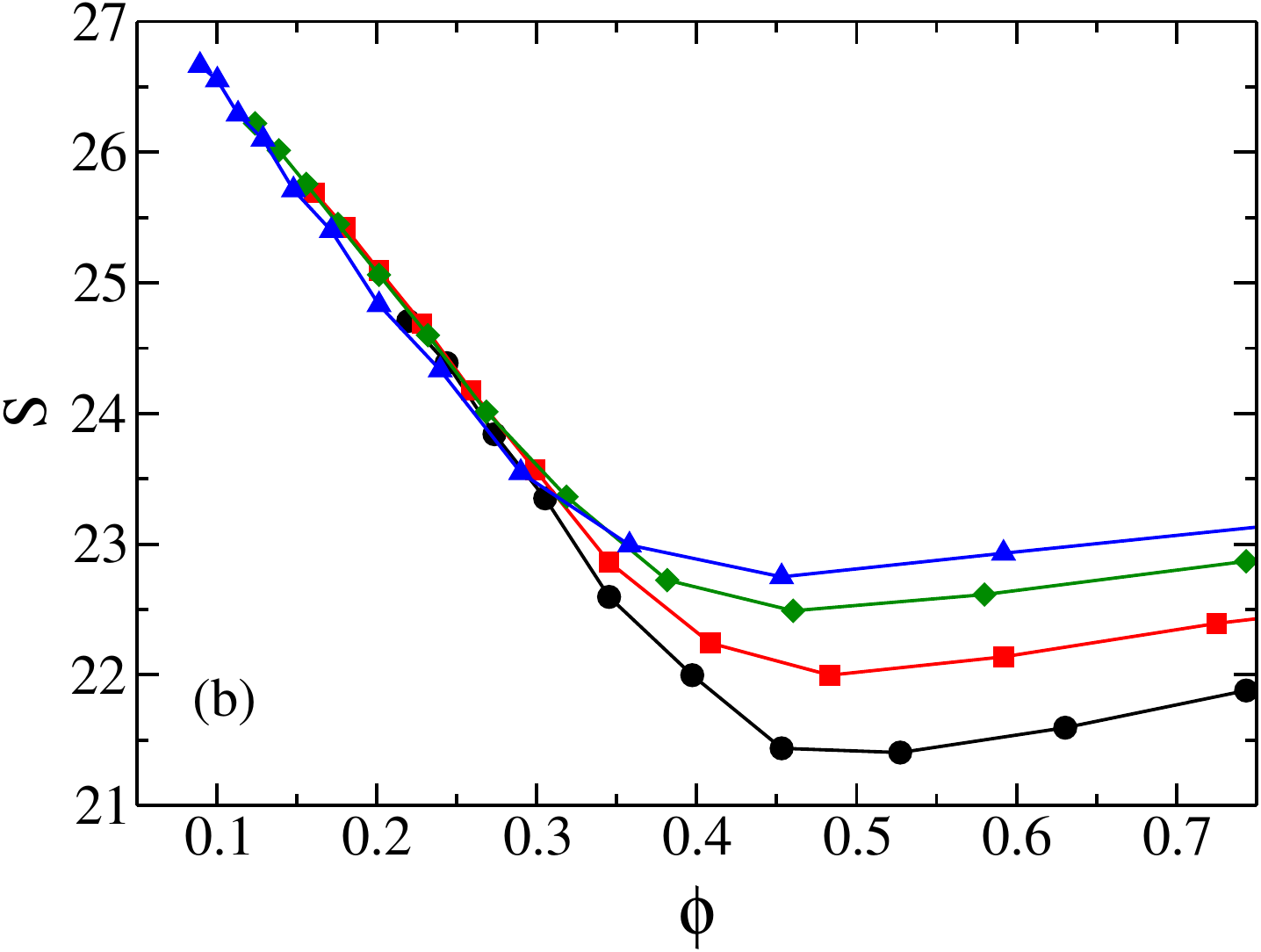}   
\caption{Figures show non-monotonic variation of entropy $S$  as a function of (a) back wall  position $X$ and b) effective space  $\phi$ for different apex angles $\alpha$. A nice collapse of data on a single line up to a certain value of $\phi$ is a apparent from the plot.}
 \label{fig2}
\end {figure}

\section{Results}
\label{l2}
\subsection{Entropy in good solvent}
While calculating entropy exactly remains a formidable challenge in most of theoretical studies, the exact enumeration technique offers a distinct advantage. With exact knowledge of the density of states, one can compute entropy by exhaustively enumerating all possible configurations within a given system's complexity. For instance, in the context of polymer translocation, exact enumeration method systematically generates  all polymer conformations within the {\it cis} and {\it trans}-side of the cone-shaped-channel providing an exact measure of configurational entropy. This approach serves as a powerful tool for understanding complex systems where analytical solutions are elusive, offering insights into the fine-grained thermodynamics that govern translocation behavior. The entropy of the system is calculated as 

\begin{equation}
 S =  k_{B} lnZ_X,
 \label{2}
\end{equation}
where $k_B$ is the Boltzmann constant.

From here on wards, we work in the reduced energy system by setting $k_B=1$. Since there is no external force on the monomer other than the confinement (due to the back-wall and conical shape of the channel), a polymer comes out from the pore as a result of the entropy difference across the pore.  Using Eq. \ref{2}, we calculate entropy associated with a polymer chain confined in the conical pore. In Fig. \ref{2}, we show the variation of entropy as a function of the position of the back wall for various apex angles. As the back wall approaches the pore, the entropy of the system decreases. This decline is attributed to the reduction in the number of configurations available to a finite-size polymer within the cone. However, beyond  a certain position, we observe a reversal, with entropy increasing once more. This may be attributed to the fact that a part of the chain moves towards the {\it trans-}side, which has more configurational space, as a result there is increase in the entropy. To emphasize this behavior, we also plot entropy as a function of the effective space $\phi$, representing the fraction of accessible sites within the cone. Remarkably, we observe a collapse of entropy onto a single curve, particularly evident up to a fraction $\phi \approx 0.3$, as shown in Fig. \ref{2}b.

\subsection{Free Energy Profile}
\begin{figure*}[t]
  \includegraphics[scale=0.6]{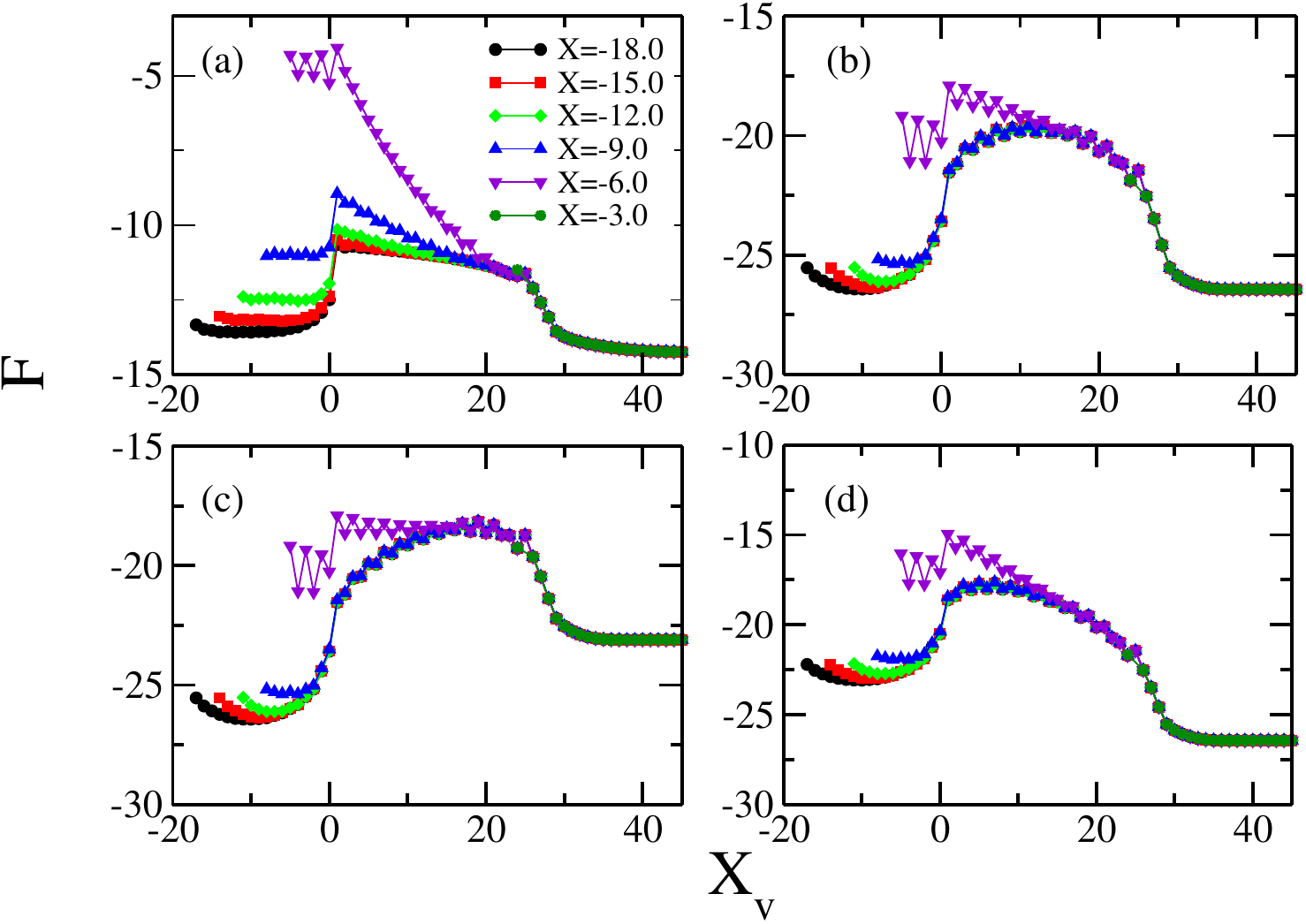}
  \caption{Free energy as a function of virtual coordinate $X_V$ (in case of apex angle $\alpha = 45^o$) for various solvent conditions: (a) good solvent on both the $\it cis$-side and $\it trans$-side ($\epsilon_c = 0.0 = \epsilon_t$), (b) poor solvent on both the $\it cis$-side and $\it trans$-side ($\epsilon_c = -1.0 = \epsilon_t$), (c) the solvent on the $\it cis$-side ($\epsilon_c=-1.0$) is poorer compared to the $\it trans-$ side ($\epsilon_t = -0.8$),  and (d) the solvent on the $\it trans$-side is poorer compared to the $\it cis$- side ($\epsilon_c = -0.8, \epsilon_t = -1.0$) }
 \label{fig3}
\end {figure*}  

A comprehensive understanding of the translocation process hinges on a complete description of the associated free energy within different regions of the channel. In Ref. \cite{Chauhan2021}, we have studied the free-energy as a function polymer-pore interaction. However, to unravel the intricate interplay between entropic confinement (arising due to the back wall position or apex angle) and translocation dynamics, we introduce variations in solvent conditions, including both good and poor solvents, across the pore. These varying solvent conditions lead to distinct polymer chain states, from swollen coils in good solvents to collapsed globules in poor solvents. To investigate polymer translocation in these diverse solvents, it is imperative to analyze the time required for each stage of the process, spanning from initiation to completion. Consequently, we divide the translocation into three pivotal stages, each holding unique significance. The initial stage involves the time taken to reach the pore, the second stage is dedicated to the threading process, and the third stage marks the full translocation to the trans-side.

Throughout these stages, the free energy is  tied to a "virtual coordinate" denoted as $X_V$, which ranges  from $-X$ to $2L$. During stage 1, when the back wall remains fixed at a specific position $X$ on the principal axis, $X_V$ corresponds to the physical distance $x$ from the pore at origin. Here, one end of the polymer is anchored at various $x$ positions on the principal axis,  while the other end is free to be any where in the  $\it cis$-side of the pore.  During this stage, $X_V$ assumes the values within the bounds of the actual coordinates $-X \le X_V < 0$.

In stage 2, the polymer commences its translocation from the $\it cis$-side to the $\it trans$-side, heralding the threading stage. Here, $X_V = m$ represents the number of monomers that have successfully translocated to the $\it trans$-side, while $(N - m - 1)$ monomers remain within the $\it cis$-side  of the pore, in addition to a monomer residing  at the pore. The final  stage 3, manifests as $X_V$ ranging from $L + 1$ to $2L$, signifying that the polymer has fully emerged from the pore and now exists entirely on the $\it trans$-side. Analogous to stage 1, here $X_V = L + |x|$, where $x$ denotes the physical distance of the polymer's anchored end on the $\it trans$-side. It is worth noting that while distances from the origin are denoted by $x$ along the real axis, the free energy is expressed here as a function of the "virtual coordinate" $X_V$. The relation between these two coordinates is given by \cite{Sun2009}

  \begin{equation}
  X_{V} =
    \begin{cases}
      - |x_{i}|, x_i= ...,-X,..,-2,-1  & \text{Stage1}\\
        m=i , i=0,1,2,3.......L & \text{Stage2}\\
       |L + x_{i}|, x_i= 1,2,...,L,...  & \text{stage3}
    \end{cases} 
    \label{3}
\end{equation}

The partition function is summed over all the walks  on a square lattice for a virtual coordinate $X_V$ with a fixed back wall position $X$ is given by the following equation:
\begin{equation}
     Z_{X}(X_V) = \sum_{N_c,N_t} C^{X_V}(N_c,N_t)u^{N_c} v^{N_t}
     \label{4}
 \end{equation} 
 where $C^{X_V}(N_c,N_t)$ is the number of conformations of respective stages as defined by Eq. \ref{3}. $N_c$ and  $N_t$ are the number of nearest neighbour contacts of the $cis$ and $trans$ sides, respectively. $u = exp(-\beta \epsilon_c)$ and $v = exp(-\beta \epsilon_t)$ are the Boltzmann weights of nearest-neighbour interactions $\epsilon_c$({\it cis}-side) and $\epsilon_t$({\it trans}-side), respectively. The  free energy of the different stages may be obtained from following equation:
\begin{equation}
     F(X_V) = -T lnZ_{X}(X_V).
     \label{5}
 \end{equation}
 $T$ is the temperature and it is set equal to 0.5.
First, we consider a situation where both {\it cis} and {\it trans} sides  have a good solvent {\it i.e.} $\epsilon_c =\epsilon_t =0$.   In a good solvent, the polymer adopts a coil-like conformation on both sides of the cone. In Fig. \ref{fig3}a, we have shown the  free-energy as a function of virtual coordinate $X_V$ for different wall positions $X$. It is evident from the plots that when one end of the polymer is anchored at a  distance far from the pore, the number of conformations of a polymer remain close  to its bulk value and the confinement by the cone-shaped channel does not affect the free-energy up to a certain value of $X_V$. However, when the anchored site shifts towards the pore, the polymer conformations get constrain by the cone and as a result its configurational properties changes relative to its bulk behavior.  Consequently, one observes a change in the  free energy at the vicinity of the pore. Once the threading process starts, the number of conformations of polymer at {\it trans-}side increases and one observes a gradual decrease in the free energy followed by a sharp fall indicating that entire polymer is translocated from the {\it cis-}side to the {\it trans-}side. As anchored end of a polymer moves away from the pore, the polymer does not experience any confinement and thus approaches its bulk behaviour.  

In the presence of a poor solvent on either side of the pore (Fig. \ref{fig3}b), we observe a slight reduction in the free energy as the anchored end of a polymer chain approaches the pore. This results due to an increase in the number of nearest neighbor contacts because of the confinement imposed by the channel. However, as the anchored site moves closer to the pore, the number of accessible conformations decreases, leading to an overall increase in the free energy. In contrast to the case of a good solvent, here during threading, the free energy remains nearly constant. One of the interesting observations is that the free energy almost remains the same for the different wall positions.  This behavior can be attributed to the polymer adopting a globule state on both sides, resulting in a sustained level of free energy. Consequently, when a monomer from the globule state on the {\it cis-}side translocates to the {\it trans-}side, it reintegrates into the globule state, thereby maintaining a constant free energy. Similar to the case with a good solvent, in the final Stage there is a sharp decrease in the free energy as the polymer assumes its bulk configuration. However, the magnitude of the free energy differs considerably compared to the good solvent.

So far we have focused on discussing the solvent quality remaining consistent across the pore. It would be interesting to observe how the free energy profile changes when the solvent quality differs across the pore.  To investigate the impact of different solvents, we consider two situations: (i) the solvent quality of {\it cis-}side is relatively poorer compared to the {\it trans-}side ($\epsilon_c < \epsilon_t$), and (ii)  the solvent quality on {\it trans-}side is relatively poorer than the {\it cis-}side ($\epsilon_c > \epsilon_t$). In Fig. \ref{fig3}c, we present the free energy profile, where the {\it cis-} and {\it trans} sides have   different interaction strengths {\it i.e. }$\epsilon_c=-1.0$  and $\epsilon_t=-0.8$. While the other features remain the same as of Fig. \ref{fig3}b, here  one observes the  free energy has an upward tilt during the threading stage. This indicates that polymer prefers to stay inside the pore.   Conversely,   Fig. \ref{fig3}d depicts the reverse situation, where the  {\it trans-}side  has the lower free energy indicating that the polymer prefers to stay out side the pore. This indicates that there is a subtle competition between entropy and the internal energy, which modifys the free energy profile. Since, the apex angle is constant ($\alpha =45^o$), the entropic contribution arising due to different conformations also remains constant. However, by varying the difference of ($\epsilon_c - \epsilon_t$), one can modify the barrier height.

\begin{figure*}[t]
  \includegraphics[scale=0.6]{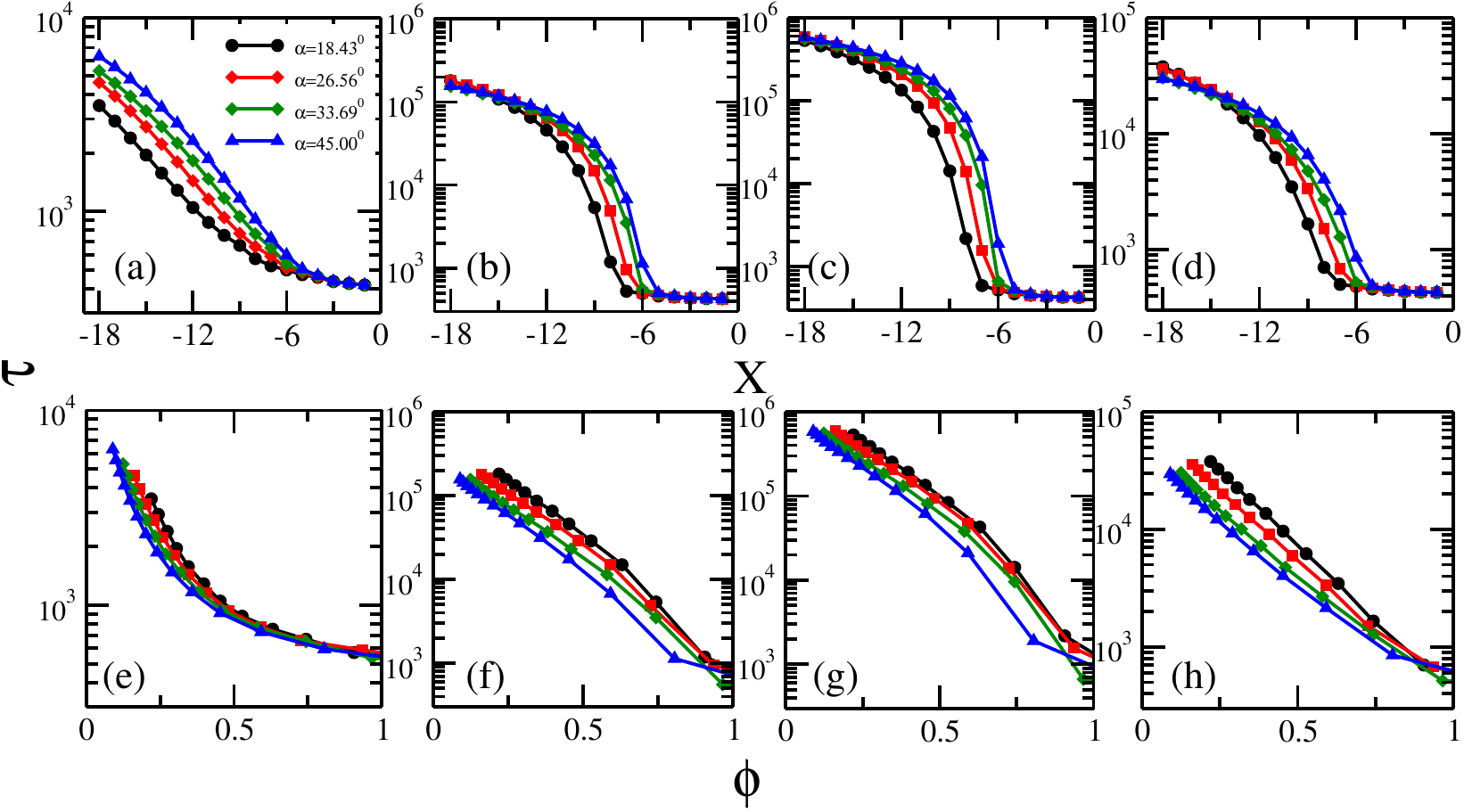}
  \caption{Figures (a-d) show the variation of translocation time as a function of the back wall position $X$, and (e-f) as a function of effective space $\phi$  for different apex angles $\alpha$. The solvent quality is varied on both the $\it cis$ and $\it trans$ side.  (a) and (e) represent the case where there is a good solvent on both sides ($\epsilon_c = 0.0 = \epsilon_t $),  (b) and (f) correspond to poor solvent conditions on either side ($\epsilon_c = -1.0 = \epsilon_t $),
  c) and (g) depict a situation where the solvent on the {\it cis-side } is poorer compared to the {\it trans-}side $\epsilon_c = -1.0, \epsilon_t = -0.8$ and, (d) and (h) present the reverse case where solvent on $\it trans$- side is poorer compared to the $\it cis$-side with $\epsilon_c = -0.8, \epsilon_t = -1.0$. }
 \label{fig4}
\end {figure*}

Another important observation is when a back wall moves toward the pore, initially the free energy of the {\it cis-}side remains almost constant, but show a significant change after $X=-12.0$. This is because that there is not enough space to accommodate a polymer at {\it cis-}side and therefore,  a significant loss of configurational entropy. Since,  a flat wall demarcates the {\it trans-side}, the number of conformations remain same irrespective of the back wall position. Therefore, the free energy after the translocation of a polymer remain almost the same for all back wall positions Figs. \ref{fig3}(a-d). When $X > -6.0$,  the space  available to the polymer inside the pore is significantly less, as a result there is a significant change in the free energy, which is apparent from Figs. \ref{fig3}(a-d). Interestingly, for $X=-3.0$, there is not enough space to accommodate even a single conformation, thus here one can see the contribution of free energy during threading and after the translocation of entire chain only. It is pertinent to mention here that the qualitative nature of the free energy profile remains the same when we change the apex angle $\alpha$. 

\subsection{Translocation Dynamics}

The phenomenon of polymer translocation can be well described by considering the average time  required for the polymer to move from the initial state ({\it cis-}side) to the final state ({\it trans-}side). This temporal parameter finds applicability in a diverse range of biological processes {\it e.g} DNA melting, polymer folding, charge hopping etc. The knowledge of this time scale provides a better understanding  of these processes at the cellular level. The diffusion of the chain can be effectively described by the Fokker-Planck equation, enabling the calculation of the time required for the chain to travel from the initial stage say $x_0$ to a specific point $x_t$, as expressed by the following equation \cite{Sung1996}

\begin{equation}
\mathcal{T}_{o \rightarrow t} = \frac{1}{D} \int_{x_{o}}^{x_{t}} e^{F(x')} dx' \int_{x_{o}}^{x'} e^{-F(x'')} dx'' .
\label{6}
\end{equation} 

\noindent The chain diffusivity, denoted as $D$, is a variable that exhibits an inverse relationship with the chain length \cite{Muthukumar2001}. In line with the approach outlined in the Ref. \cite{Sun2009}, we assign a value of $D = 1$. 
Given the implementation of a lattice model, Eq. \ref{6} can be represented in a discrete form as 

\begin{equation}
\mathcal{T}_{o \rightarrow t} = \sum_{x'=x_{o}}^{x_{t}} e^{F(x')} \Delta x' \sum_{x''=x_{o}}^{x'} e^{-F(x'')} \Delta x'' ,
\label{7}
\end{equation}

\noindent where $\Delta x'= \Delta x''= 1$ because free energy is calculated at equi-spaced lattice.
 The time ($\tau$) required for the  polymer to cross the energy barrier imparted by the pore is highly dependent on the nature of pore geometry and solvent present on either sides of the pore. We first consider a good solvent on both side of the pore (($\epsilon_c = \epsilon_t=0$)) where there is no attractive interaction among monomers.  The Figs. \ref{fig4}(a) \& (e) show the decay of translocation time ($\tau$)  for various angles as a function of  back wall position $X$ and $\phi$, respectively.  One can notice that for the apex angle $\alpha =45^o$,  the value of $\tau$ is higher compared to $\alpha=18.43^o$ for the lower values of $X$ and there is a significant difference in $\tau$ for different values of $\alpha$. Thus, one can infer that confinement promotes the translocation. However, when the back wall is closed to pore ($X>-6.0$), the value of $\tau$ tends to  a constant value.  It is interesting to note that when we plot variation of $\tau$  as a function of $\phi$, all the data collapse nearly on a single curve indicating that for the good solvent, the translocation time $\tau$ depends only on the fraction of space available to a polymer chain.

If the solvent quality is  poor across the pore($\epsilon_c = \epsilon_t=-1.0$), we find that the translocation time decay  up to a certain value of $X$ and then approaches to a constant value (Fig. \ref{4}(b)). In this case, polymer acquires the conformation of globule state of relatively smaller size compared to the swollen state.  Therefore, if the wall is far away from the pore, translocation time remains constant for all apex angles studied here as polymer does not see the confinement. As the back wall approaches towards the pore, we observe a clear difference in $\tau$. Like the good solvent case, we also find that confinement advances the translocation, however the value of $\tau$ is almost an order higher than the one observed for the good solvent. The rise in the $\tau$ may be understood from the Fig. \ref{fig4}(b), where we observed that the free energy remains constant during threading. Unlike the good solvent, in this case the translocation times are well separated and show a linear decrease with $\phi$ for all angles. For  $\phi =1$, the translocation times for all angles tend to be nearly equal as there is no space in the $\it cis-$side and the  $\it trans-$side has equal space for all values of $\alpha$.

Interestingly, when the solvent quality across the pore are different ($\epsilon_c=-1.0$ $\epsilon_t=-0.8$) and  ($\epsilon_c=-0.8$ $\epsilon_t=-1$), we  do not see any qualitative change in the translocation behaviour except the value of $\tau$.  If the $\it cis-$side has relatively poorer solvent, the free energy during threading has up-hill trends and thus hinders the polymer's ability to thread through the pore,  resulting in increased translocation times compared to the similar solvent ($\epsilon_c = \epsilon_t=-1.0$) across the pore. Finally, for  the reverse solvent conditions,  ($\epsilon_c=-0.8$ and $\epsilon_t=-1.0$), the free energy has  down-hill trends during the threading and hence polymer would prefer to be on the $\it trans-$side resulting a net decrease in $\tau$.

\begin{figure}[t]
  \includegraphics[scale=0.3]{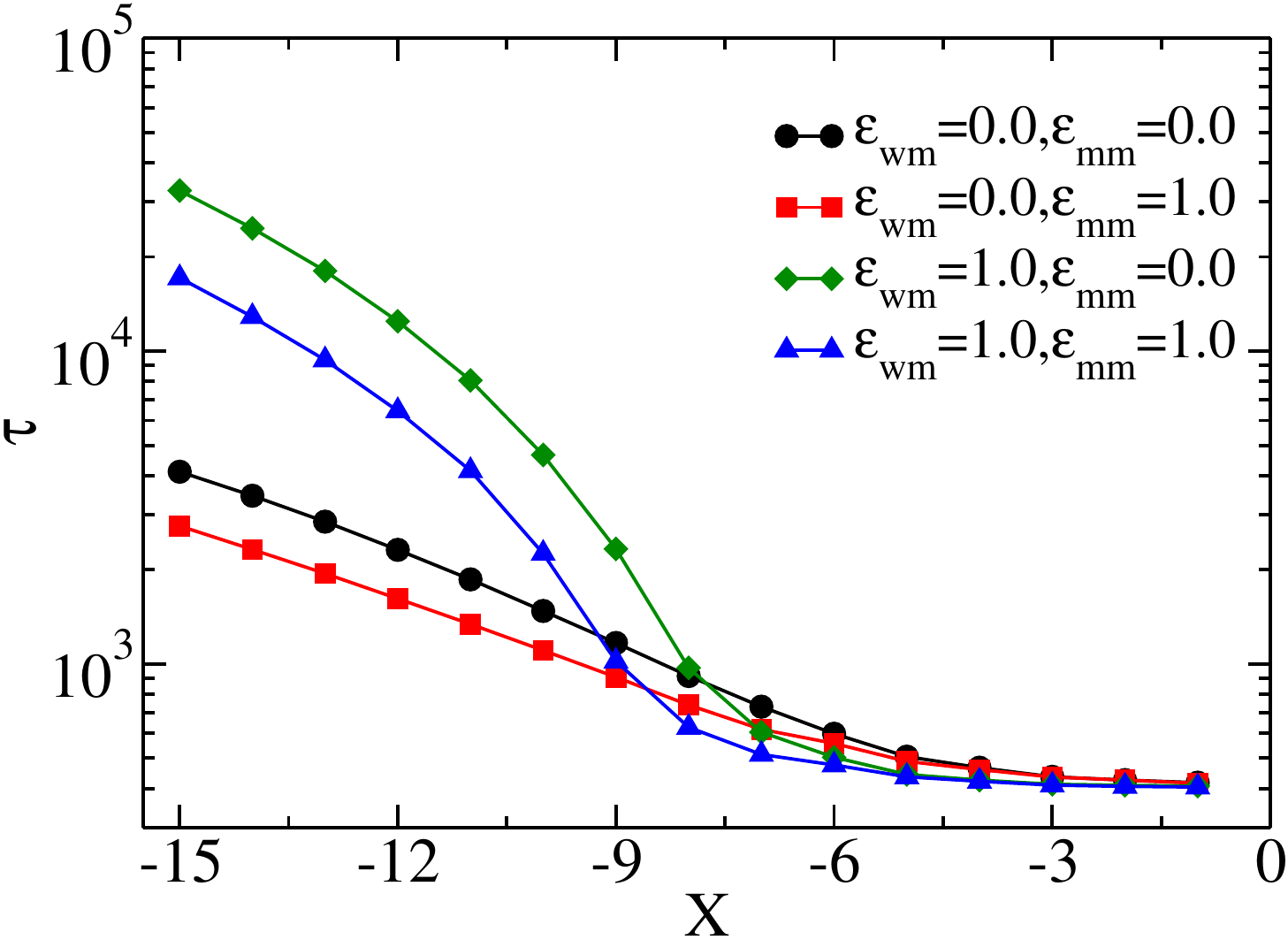}
  \caption{The variation of translocation time as a function of the back wall position $X$. Four distinct cases are investigated. $\epsilon_{mm}$ and $\epsilon_{wm}$ are the interactions among monomers and between monomers and channel walls respectively.}
 \label{fig5}
\end {figure}  

\section{Influence of Repulsive Forces Among Monomers and Pore Wall in Polymer Translocation }
\label{l3}
We now extend our study where the pore wall is either repulsive or neutral for a polymer. If we consider repulsive interaction in addition to excluded volume in polymer, the model may mimic the case of charged DNA. For this we choose apex angle $\alpha =45^o$ and vary the back-wall position. We systematically investigate the intricate dynamics of polymer translocation for four distinct cases, shown in Fig. \ref{fig5}, each offering insights into the specific molecular interactions that play a crucial role in the translocation of charged DNA through confined geometries. Firstly, we show the results where no interactions were considered among  monomers, and between monomers and the channel walls ($\epsilon_{mm}=0.0$ and $\epsilon_{wm}=0.0$) to establish a fundamental baseline scenario of translocation dynamics.  This setting provides a crucial reference point against which subsequent cases could be compared. Next, we consider a scenario with no monomer-wall interaction but included a repulsive force among monomers ($\epsilon_{wm}=0.0$ and $\epsilon_{mm}=1.0$). To our surprise, this configuration resulted in a minimization of translocation time. This may be understood as the repulsive interaction among monomers serves to prevent entanglements (end-to end distance increases), reducing the likelihood of collisions during the translocation process. In the absence of interactions with the channel walls, monomers experience enhanced freedom of movement, allowing for a smoother and more efficient passage through the cone-shaped channel. This reduced hindrance to motion, facilitated by repulsive forces among monomers, contributes to their efficient navigation within the channel. Moreover, the repulsive interactions foster an optimal exploration of the available space, minimizing unnecessary delays and promoting a streamlined progression through the channel.

Interestingly, in the scenario where the only repulsive interaction existed between the monomer and the pore wall, excluding interactions among monomers ($\epsilon_{wm}=1.0$ and $\epsilon_{mm}=0.0$), we observed an unexpected increase in translocation time compared to the non-interaction scenario. This unexpected delay can be attributed to the influential role of the repulsive forces in the cone-shaped channel, leading to reduced available space for the polymer. This confinement effect results in a reduction in polymer size due to increased repulsive interaction with the channel walls. Consequently, we observed a delay in translocation time when the monomer-wall interaction was in play, highlighting the significant impact of the repulsive forces in the cone-shaped channel on the translocation dynamics.

The investigation into the combined effects of monomer-monomer and monomer-wall repulsive interactions ($\epsilon_{mm}=1.0$ and $\epsilon_{wm}=1.0$) revealed intriguing dynamics during translocation. When both types of repulsive forces are in play, the polymer encounters a nuanced environment that influences its movement through the cone-shaped channel. The repulsive interaction among monomers prevents excessive clustering, minimizing collisions and entanglements during translocation. Simultaneously, the repulsive force between monomers and the channel wall adds another layer of complexity. The channel, by nature, constrains the available space for the polymer, leading to a reduction in polymer size due to confinement, ultimately affecting the end-to-end distance. This reduction in available space and polymer size contributes to a delayed translocation time compared to scenarios with no interactions among monomers and walls. However, intriguingly, the translocation in this combined scenario is faster than when only the wall exerts a repulsive force. This suggests a delicate interplay between various forces, emphasizing the importance of understanding the nuanced effects of multiple interactions in the translocation process.

These findings not only contribute to our understanding of polymer translocation dynamics through cone-shaped channels but also offer valuable insights into the specific challenges and optimizations associated with the translocation of polymer through confined geometries. The unexpected trends observed in certain scenarios open avenues for further exploration and the potential refinement of translocation processes, particularly in the context of charged DNA.

\section{conclusions}
\label{l4}
In this study, we have investigated the influence of confinement on  polymer translocation through a cone-shaped pore using the exact enumeration technique.  In contrast to previous investigations, we introduce a finite space on the {\it cis-}side that can be varied, while the {\it trans}-side remains semi-infinite. The space available at {\it cis-side} can be varied by placing a back wall at different positions away from the pore. Additionally, the confinement is also varied by changing the apex angle $\alpha$, allowing us to observe its effect on polymer translocation. One end of the polymer is  anchored on  the principal axis, while the other end is free to be anywhere. We systematically vary the position of anchored sites from  the {\it cis-}side to the {\it trans-}side, and obtain the exact free energy of the system for a given value of back wall position $X$ and the apex angle of the pore $\alpha$. Using the Fokker-Planck equation, we obtained translocation times from the free energy of the system having different solvent qualities across the pore.

We place particular emphasis on the novel contribution made in this work is the calculation of system entropy exactly, which adds the understanding to the translocation process. Our results, depicted in Fig. \ref{2}(a)), reveal a clear trend: as the back wall position $X$ shifts, the system's entropy decreases. What's more, this effect is amplified when the apex angle $\alpha$ is reduced. Strikingly, when plotting entropy as a function of $\phi$ for all $\alpha$ values, interestingly all the data points collapse onto a single line. This intriguing finding suggests that, up to a certain $\phi$ threshold, entropy primarily hinges on the space accessible to the polymer within the {\it cis}-side. As the anchored site approaches the pore, the system's entropy rises, influenced by the portion of the chain that translocates to the trans-side, boasting semi-infinite space. This intriguing phenomenon ultimately leads to the system's entropy reaching a constant value, irrespective of values of $\alpha$.

The translocation time ($\tau$) obtained from the free energy analysis provides valuable insights into the translocation process within a confined environment. In the case of a good solvent, $\tau$ decreases as the level of confinement increases either due to the position of the back wall or the confining angle. Remarkably, the most significant finding is that the values of $\tau$  collapse onto a single curve when plotted against $\phi$. This suggests that the translocation time is primarily influenced by the confining space, regardless of its specific shape. Based on this observation, we propose that the inclusion of molecular crowders or similar entities within the pore can potentially expedite the translocation process.

Another interesting finding is that by manipulating the difference in solvent quality across the pore, one can exert control over the translocation process. The linear decrease in $\tau$ with $\phi$ for different values of $\alpha$ indicates the need for further investigation into scaling phenomena. This feature holds potential applications in various fields such as cell metabolism, DNA-RNA sorting and sequencing, and drug delivery mechanisms. These applications involve the transport of biomolecules, which is influenced by the chemical potential gradient of the solvent. We hope that future measurements of translocation times under variable confining environments will contribute to a better understanding of the translocation process.

We also explored polymer translocation through a cone-shaped channel with either repulsive or neutral interactions with the pore wall, mimicking scenarios relevant to charged DNA translocation. Investigating four distinct cases, we uncover intricate dynamics shaping the translocation of charged DNA through confined geometries. The absence of interactions among monomers and channel walls establishes a fundamental baseline, while the introduction of repulsive forces among monomers surprisingly minimizes translocation time, preventing entanglements and enhancing freedom of movement. Unexpected delays arise when only the pore wall exerts a repulsive force, reducing available space and polymer size. Combined monomer-monomer and monomer-wall repulsive interactions introduce nuanced dynamics, emphasizing the intricate interplay between forces and contributing to a delayed but faster translocation compared to scenarios with only wall repulsion. This study provides comprehensive insights into the complex factors influencing polymer translocation through confined channels.

Financial assistance from the SERB, India, UGC, India, SPARC scheme of MoE, and IoE scheme, MoE, India are gratefully acknowledged.

\bibliography{main}

\begin{thebibliography}{42}%
\makeatletter
\providecommand \@ifxundefined [1]{%
 \@ifx{#1\undefined}
}%
\providecommand \@ifnum [1]{%
 \ifnum #1\expandafter \@firstoftwo
 \else \expandafter \@secondoftwo
 \fi
}%
\providecommand \@ifx [1]{%
 \ifx #1\expandafter \@firstoftwo
 \else \expandafter \@secondoftwo
 \fi
}%
\providecommand \natexlab [1]{#1}%
\providecommand \enquote  [1]{``#1''}%
\providecommand \bibnamefont  [1]{#1}%
\providecommand \bibfnamefont [1]{#1}%
\providecommand \citenamefont [1]{#1}%
\providecommand \href@noop [0]{\@secondoftwo}%
\providecommand \href [0]{\begingroup \@sanitize@url \@href}%
\providecommand \@href[1]{\@@startlink{#1}\@@href}%
\providecommand \@@href[1]{\endgroup#1\@@endlink}%
\providecommand \@sanitize@url [0]{\catcode `\\12\catcode `\$12\catcode
  `\&12\catcode `\#12\catcode `\^12\catcode `\_12\catcode `\%12\relax}%
\providecommand \@@startlink[1]{}%
\providecommand \@@endlink[0]{}%
\providecommand \url  [0]{\begingroup\@sanitize@url \@url }%
\providecommand \@url [1]{\endgroup\@href {#1}{\urlprefix }}%
\providecommand \urlprefix  [0]{URL }%
\providecommand \Eprint [0]{\href }%
\providecommand \doibase [0]{http://dx.doi.org/}%
\providecommand \selectlanguage [0]{\@gobble}%
\providecommand \bibinfo  [0]{\@secondoftwo}%
\providecommand \bibfield  [0]{\@secondoftwo}%
\providecommand \translation [1]{[#1]}%
\providecommand \BibitemOpen [0]{}%
\providecommand \bibitemStop [0]{}%
\providecommand \bibitemNoStop [0]{.\EOS\space}%
\providecommand \EOS [0]{\spacefactor3000\relax}%
\providecommand \BibitemShut  [1]{\csname bibitem#1\endcsname}%
\let\auto@bib@innerbib\@empty
\bibitem [{\citenamefont {Vella}(1994)}]{Vella1994}%
  \BibitemOpen
  \bibfield  {author} {\bibinfo {author} {\bibfnamefont {F.}~\bibnamefont
  {Vella}},\ }\href {https://doi.org/10.1016/0307-4412(94)90059-0} {\bibfield
  {journal} {\bibinfo  {journal} {Biochemical Education}\ }\textbf {\bibinfo
  {volume} {22}},\ \bibinfo {pages} {164} (\bibinfo {year} {1994})}\BibitemShut
  {NoStop}%
\bibitem [{\citenamefont {Nakielny}\ and\ \citenamefont
  {Dreyfuss}(1999)}]{Nakielny1999}%
  \BibitemOpen
  \bibfield  {author} {\bibinfo {author} {\bibfnamefont {S.}~\bibnamefont
  {Nakielny}}\ and\ \bibinfo {author} {\bibfnamefont {G.}~\bibnamefont
  {Dreyfuss}},\ }\href {https://doi.org/10.1016/s0092-8674(00)81666-9}
  {\bibfield  {journal} {\bibinfo  {journal} {Cell}\ }\textbf {\bibinfo
  {volume} {99}},\ \bibinfo {pages} {677} (\bibinfo {year} {1999})}\BibitemShut
  {NoStop}%
\bibitem [{\citenamefont {Molineux}\ and\ \citenamefont
  {Panja}(2013)}]{Molineux2013}%
  \BibitemOpen
  \bibfield  {author} {\bibinfo {author} {\bibfnamefont {I.~J.}\ \bibnamefont
  {Molineux}}\ and\ \bibinfo {author} {\bibfnamefont {D.}~\bibnamefont
  {Panja}},\ }\href {https://doi.org/10.1038/nrmicro2988} {\bibfield  {journal}
  {\bibinfo  {journal} {Nature Reviews Microbiology}\ }\textbf {\bibinfo
  {volume} {11}},\ \bibinfo {pages} {194} (\bibinfo {year} {2013})}\BibitemShut
  {NoStop}%
\bibitem [{\citenamefont {Marenduzzo}\ \emph {et~al.}(2013)\citenamefont
  {Marenduzzo}, \citenamefont {Micheletti}, \citenamefont {Orlandini},\ and\
  \citenamefont {Sumners}}]{Marenduzzo2013}%
  \BibitemOpen
  \bibfield  {author} {\bibinfo {author} {\bibfnamefont {D.}~\bibnamefont
  {Marenduzzo}}, \bibinfo {author} {\bibfnamefont {C.}~\bibnamefont
  {Micheletti}}, \bibinfo {author} {\bibfnamefont {E.}~\bibnamefont
  {Orlandini}}, \ and\ \bibinfo {author} {\bibfnamefont {D.~W.}\ \bibnamefont
  {Sumners}},\ }\href {https://doi.org/10.1073/pnas.1306601110} {\bibfield
  {journal} {\bibinfo  {journal} {Proceedings of the National Academy of
  Sciences}\ }\textbf {\bibinfo {volume} {110}},\ \bibinfo {pages} {20081}
  (\bibinfo {year} {2013})}\BibitemShut {NoStop}%
\bibitem [{\citenamefont {Berndsen}\ \emph {et~al.}(2014)\citenamefont
  {Berndsen}, \citenamefont {Keller}, \citenamefont {Grimes}, \citenamefont
  {Jardine},\ and\ \citenamefont {Smith}}]{Berndsen2014}%
  \BibitemOpen
  \bibfield  {author} {\bibinfo {author} {\bibfnamefont {Z.~T.}\ \bibnamefont
  {Berndsen}}, \bibinfo {author} {\bibfnamefont {N.}~\bibnamefont {Keller}},
  \bibinfo {author} {\bibfnamefont {S.}~\bibnamefont {Grimes}}, \bibinfo
  {author} {\bibfnamefont {P.~J.}\ \bibnamefont {Jardine}}, \ and\ \bibinfo
  {author} {\bibfnamefont {D.~E.}\ \bibnamefont {Smith}},\ }\href
  {https://doi.org/10.1073/pnas.1405109111} {\bibfield  {journal} {\bibinfo
  {journal} {Proceedings of the National Academy of Sciences}\ }\textbf
  {\bibinfo {volume} {111}},\ \bibinfo {pages} {8345} (\bibinfo {year}
  {2014})}\BibitemShut {NoStop}%
\bibitem [{\citenamefont {Kasianowicz}\ \emph {et~al.}(1996)\citenamefont
  {Kasianowicz}, \citenamefont {Brandin}, \citenamefont {Branton},\ and\
  \citenamefont {Deamer}}]{Kasianowicz1996}%
  \BibitemOpen
  \bibfield  {author} {\bibinfo {author} {\bibfnamefont {J.}~\bibnamefont
  {Kasianowicz}}, \bibinfo {author} {\bibfnamefont {E.}~\bibnamefont
  {Brandin}}, \bibinfo {author} {\bibfnamefont {D.}~\bibnamefont {Branton}}, \
  and\ \bibinfo {author} {\bibfnamefont {D.}~\bibnamefont {Deamer}},\ }\href
  {https://doi.org/10.1073/pnas.93.24.13770} {\bibfield  {journal} {\bibinfo
  {journal} {Proceedings of the National Academy of Sciences}\ }\textbf
  {\bibinfo {volume} {93}},\ \bibinfo {pages} {13770} (\bibinfo {year}
  {1996})}\BibitemShut {NoStop}%
\bibitem [{\citenamefont {Schneider}\ \emph {et~al.}(2010)\citenamefont
  {Schneider}, \citenamefont {Kowalczyk}, \citenamefont {Calado}, \citenamefont
  {Pandraud}, \citenamefont {Zandbergen}, \citenamefont {Vandersypen},\ and\
  \citenamefont {Dekker}}]{Schneider2010}%
  \BibitemOpen
  \bibfield  {author} {\bibinfo {author} {\bibfnamefont {G.~F.}\ \bibnamefont
  {Schneider}}, \bibinfo {author} {\bibfnamefont {S.~W.}\ \bibnamefont
  {Kowalczyk}}, \bibinfo {author} {\bibfnamefont {V.~E.}\ \bibnamefont
  {Calado}}, \bibinfo {author} {\bibfnamefont {G.}~\bibnamefont {Pandraud}},
  \bibinfo {author} {\bibfnamefont {H.~W.}\ \bibnamefont {Zandbergen}},
  \bibinfo {author} {\bibfnamefont {L.~M.~K.}\ \bibnamefont {Vandersypen}}, \
  and\ \bibinfo {author} {\bibfnamefont {C.}~\bibnamefont {Dekker}},\ }\href
  {https://doi.org/10.1021/nl102069z} {\bibfield  {journal} {\bibinfo
  {journal} {Nano Letters}\ }\textbf {\bibinfo {volume} {10}},\ \bibinfo
  {pages} {3163} (\bibinfo {year} {2010})}\BibitemShut {NoStop}%
\bibitem [{\citenamefont {Luo}\ \emph {et~al.}(2007)\citenamefont {Luo},
  \citenamefont {Ala-Nissila}, \citenamefont {Ying},\ and\ \citenamefont
  {Bhattacharya}}]{Luo2007}%
  \BibitemOpen
  \bibfield  {author} {\bibinfo {author} {\bibfnamefont {K.}~\bibnamefont
  {Luo}}, \bibinfo {author} {\bibfnamefont {T.}~\bibnamefont {Ala-Nissila}},
  \bibinfo {author} {\bibfnamefont {S.-C.}\ \bibnamefont {Ying}}, \ and\
  \bibinfo {author} {\bibfnamefont {A.}~\bibnamefont {Bhattacharya}},\ }\href
  {https://doi.org/10.1103/physrevlett.99.148102} {\bibfield  {journal}
  {\bibinfo  {journal} {Physical Review Letters}\ }\textbf {\bibinfo {volume}
  {99}} (\bibinfo {year} {2007})}\BibitemShut {NoStop}%
\bibitem [{\citenamefont {Chen}\ \emph {et~al.}(2009)\citenamefont {Chen},
  \citenamefont {Wang}, \citenamefont {Zhou},\ and\ \citenamefont
  {Luo}}]{Chen2009}%
  \BibitemOpen
  \bibfield  {author} {\bibinfo {author} {\bibfnamefont {Y.-C.}\ \bibnamefont
  {Chen}}, \bibinfo {author} {\bibfnamefont {C.}~\bibnamefont {Wang}}, \bibinfo
  {author} {\bibfnamefont {Y.-L.}\ \bibnamefont {Zhou}}, \ and\ \bibinfo
  {author} {\bibfnamefont {M.-B.}\ \bibnamefont {Luo}},\ }\href
  {https://doi.org/10.1063/1.3071198} {\bibfield  {journal} {\bibinfo
  {journal} {The Journal of Chemical Physics}\ }\textbf {\bibinfo {volume}
  {130}},\ \bibinfo {pages} {054902} (\bibinfo {year} {2009})}\BibitemShut
  {NoStop}%
\bibitem [{\citenamefont {Luo}\ and\ \citenamefont {Cao}(2012)}]{Luo2012}%
  \BibitemOpen
  \bibfield  {author} {\bibinfo {author} {\bibfnamefont {M.-B.}\ \bibnamefont
  {Luo}}\ and\ \bibinfo {author} {\bibfnamefont {W.-P.}\ \bibnamefont {Cao}},\
  }\href {https://doi.org/10.1103/physreve.86.031914} {\bibfield  {journal}
  {\bibinfo  {journal} {Physical Review E}\ }\textbf {\bibinfo {volume} {86}}
  (\bibinfo {year} {2012})}\BibitemShut {NoStop}%
\bibitem [{\citenamefont {Polson}\ and\ \citenamefont
  {Heckbert}(2019)}]{Polson2019}%
  \BibitemOpen
  \bibfield  {author} {\bibinfo {author} {\bibfnamefont {J.~M.}\ \bibnamefont
  {Polson}}\ and\ \bibinfo {author} {\bibfnamefont {D.~R.}\ \bibnamefont
  {Heckbert}},\ }\href {https://doi.org/10.1103/physreve.100.012504} {\bibfield
   {journal} {\bibinfo  {journal} {Physical Review E}\ }\textbf {\bibinfo
  {volume} {100}} (\bibinfo {year} {2019})}\BibitemShut {NoStop}%
\bibitem [{\citenamefont {Mohan}\ \emph {et~al.}(2010)\citenamefont {Mohan},
  \citenamefont {Kolomeisky},\ and\ \citenamefont {Pasquali}}]{Mohan2010}%
  \BibitemOpen
  \bibfield  {author} {\bibinfo {author} {\bibfnamefont {A.}~\bibnamefont
  {Mohan}}, \bibinfo {author} {\bibfnamefont {A.~B.}\ \bibnamefont
  {Kolomeisky}}, \ and\ \bibinfo {author} {\bibfnamefont {M.}~\bibnamefont
  {Pasquali}},\ }\href {https://doi.org/10.1063/1.3458821} {\bibfield
  {journal} {\bibinfo  {journal} {The Journal of Chemical Physics}\ }\textbf
  {\bibinfo {volume} {133}},\ \bibinfo {pages} {024902} (\bibinfo {year}
  {2010})}\BibitemShut {NoStop}%
\bibitem [{\citenamefont {Luo}\ \emph {et~al.}(2006)\citenamefont {Luo},
  \citenamefont {Huopaniemi}, \citenamefont {Ala-Nissila},\ and\ \citenamefont
  {Ying}}]{Luo2006}%
  \BibitemOpen
  \bibfield  {author} {\bibinfo {author} {\bibfnamefont {K.}~\bibnamefont
  {Luo}}, \bibinfo {author} {\bibfnamefont {I.}~\bibnamefont {Huopaniemi}},
  \bibinfo {author} {\bibfnamefont {T.}~\bibnamefont {Ala-Nissila}}, \ and\
  \bibinfo {author} {\bibfnamefont {S.-C.}\ \bibnamefont {Ying}},\ }\href
  {https://doi.org/10.1063/1.2179792} {\bibfield  {journal} {\bibinfo
  {journal} {The Journal of Chemical Physics}\ }\textbf {\bibinfo {volume}
  {124}},\ \bibinfo {pages} {114704} (\bibinfo {year} {2006})}\BibitemShut
  {NoStop}%
\bibitem [{\citenamefont {Dabhade}\ \emph {et~al.}(2022)\citenamefont
  {Dabhade}, \citenamefont {Chauhan},\ and\ \citenamefont
  {Chaudhury}}]{Dabhade2022}%
  \BibitemOpen
  \bibfield  {author} {\bibinfo {author} {\bibfnamefont {A.}~\bibnamefont
  {Dabhade}}, \bibinfo {author} {\bibfnamefont {A.}~\bibnamefont {Chauhan}}, \
  and\ \bibinfo {author} {\bibfnamefont {S.}~\bibnamefont {Chaudhury}},\ }\href
  {https://doi.org/10.1002/cphc.202200666} {\bibfield  {journal} {\bibinfo
  {journal} {{ChemPhysChem}}\ }\textbf {\bibinfo {volume} {24}} (\bibinfo
  {year} {2022})}\BibitemShut {NoStop}%
\bibitem [{\citenamefont {Hsiao}(2020)}]{Hsiao2020}%
  \BibitemOpen
  \bibfield  {author} {\bibinfo {author} {\bibfnamefont {P.-Y.}\ \bibnamefont
  {Hsiao}},\ }\href {https://doi.org/10.1021/acsomega.0c02647} {\bibfield
  {journal} {\bibinfo  {journal} {{ACS} Omega}\ }\textbf {\bibinfo {volume}
  {5}},\ \bibinfo {pages} {19805} (\bibinfo {year} {2020})}\BibitemShut
  {NoStop}%
\bibitem [{\citenamefont {Luo}\ \emph {et~al.}(2008)\citenamefont {Luo},
  \citenamefont {Ollila}, \citenamefont {Huopaniemi}, \citenamefont
  {Ala-Nissila}, \citenamefont {Pomorski}, \citenamefont {Karttunen},
  \citenamefont {Ying},\ and\ \citenamefont {Bhattacharya}}]{Luo2008}%
  \BibitemOpen
  \bibfield  {author} {\bibinfo {author} {\bibfnamefont {K.}~\bibnamefont
  {Luo}}, \bibinfo {author} {\bibfnamefont {S.~T.~T.}\ \bibnamefont {Ollila}},
  \bibinfo {author} {\bibfnamefont {I.}~\bibnamefont {Huopaniemi}}, \bibinfo
  {author} {\bibfnamefont {T.}~\bibnamefont {Ala-Nissila}}, \bibinfo {author}
  {\bibfnamefont {P.}~\bibnamefont {Pomorski}}, \bibinfo {author}
  {\bibfnamefont {M.}~\bibnamefont {Karttunen}}, \bibinfo {author}
  {\bibfnamefont {S.-C.}\ \bibnamefont {Ying}}, \ and\ \bibinfo {author}
  {\bibfnamefont {A.}~\bibnamefont {Bhattacharya}},\ }\href
  {https://doi.org/10.1103/physreve.78.050901} {\bibfield  {journal} {\bibinfo
  {journal} {Physical Review E}\ }\textbf {\bibinfo {volume} {78}} (\bibinfo
  {year} {2008})}\BibitemShut {NoStop}%
\bibitem [{\citenamefont {Edmonds}\ \emph {et~al.}(2012)\citenamefont
  {Edmonds}, \citenamefont {Hudiono}, \citenamefont {Ahmadi}, \citenamefont
  {Hesketh},\ and\ \citenamefont {Nair}}]{Edmonds2012}%
  \BibitemOpen
  \bibfield  {author} {\bibinfo {author} {\bibfnamefont {C.~M.}\ \bibnamefont
  {Edmonds}}, \bibinfo {author} {\bibfnamefont {Y.~C.}\ \bibnamefont
  {Hudiono}}, \bibinfo {author} {\bibfnamefont {A.~G.}\ \bibnamefont {Ahmadi}},
  \bibinfo {author} {\bibfnamefont {P.~J.}\ \bibnamefont {Hesketh}}, \ and\
  \bibinfo {author} {\bibfnamefont {S.}~\bibnamefont {Nair}},\ }\href
  {https://doi.org/10.1063/1.3682777} {\bibfield  {journal} {\bibinfo
  {journal} {The Journal of Chemical Physics}\ }\textbf {\bibinfo {volume}
  {136}},\ \bibinfo {pages} {065105} (\bibinfo {year} {2012})}\BibitemShut
  {NoStop}%
\bibitem [{\citenamefont {Ikonen}\ \emph {et~al.}(2012)\citenamefont {Ikonen},
  \citenamefont {Bhattacharya}, \citenamefont {Ala-Nissila},\ and\
  \citenamefont {Sung}}]{Ikonen2012}%
  \BibitemOpen
  \bibfield  {author} {\bibinfo {author} {\bibfnamefont {T.}~\bibnamefont
  {Ikonen}}, \bibinfo {author} {\bibfnamefont {A.}~\bibnamefont
  {Bhattacharya}}, \bibinfo {author} {\bibfnamefont {T.}~\bibnamefont
  {Ala-Nissila}}, \ and\ \bibinfo {author} {\bibfnamefont {W.}~\bibnamefont
  {Sung}},\ }\href {https://doi.org/10.1063/1.4742188} {\bibfield  {journal}
  {\bibinfo  {journal} {The Journal of Chemical Physics}\ }\textbf {\bibinfo
  {volume} {137}},\ \bibinfo {pages} {085101} (\bibinfo {year}
  {2012})}\BibitemShut {NoStop}%
\bibitem [{\citenamefont {Bhattacharya}\ \emph {et~al.}(2009)\citenamefont
  {Bhattacharya}, \citenamefont {Morrison}, \citenamefont {Luo}, \citenamefont
  {Ala-Nissila}, \citenamefont {Ying}, \citenamefont {Milchev},\ and\
  \citenamefont {Binder}}]{Bhattacharya2009}%
  \BibitemOpen
  \bibfield  {author} {\bibinfo {author} {\bibfnamefont {A.}~\bibnamefont
  {Bhattacharya}}, \bibinfo {author} {\bibfnamefont {W.~H.}\ \bibnamefont
  {Morrison}}, \bibinfo {author} {\bibfnamefont {K.}~\bibnamefont {Luo}},
  \bibinfo {author} {\bibfnamefont {T.}~\bibnamefont {Ala-Nissila}}, \bibinfo
  {author} {\bibfnamefont {S.~C.}\ \bibnamefont {Ying}}, \bibinfo {author}
  {\bibfnamefont {A.}~\bibnamefont {Milchev}}, \ and\ \bibinfo {author}
  {\bibfnamefont {K.}~\bibnamefont {Binder}},\ }\href
  {https://doi.org/10.1140/epje/i2009-10495-5} {\bibfield  {journal} {\bibinfo
  {journal} {The European Physical Journal E}\ }\textbf {\bibinfo {volume}
  {29}},\ \bibinfo {pages} {423} (\bibinfo {year} {2009})}\BibitemShut
  {NoStop}%
\bibitem [{\citenamefont {Abdolvahab}\ \emph {et~al.}(2011)\citenamefont
  {Abdolvahab}, \citenamefont {Ejtehadi},\ and\ \citenamefont
  {Metzler}}]{Abdolvahab2011}%
  \BibitemOpen
  \bibfield  {author} {\bibinfo {author} {\bibfnamefont {R.~H.}\ \bibnamefont
  {Abdolvahab}}, \bibinfo {author} {\bibfnamefont {M.~R.}\ \bibnamefont
  {Ejtehadi}}, \ and\ \bibinfo {author} {\bibfnamefont {R.}~\bibnamefont
  {Metzler}},\ }\href {https://doi.org/10.1103/physreve.83.011902} {\bibfield
  {journal} {\bibinfo  {journal} {Physical Review E}\ }\textbf {\bibinfo
  {volume} {83}} (\bibinfo {year} {2011})}\BibitemShut {NoStop}%
\bibitem [{\citenamefont {Suhonen}\ and\ \citenamefont
  {Linna}(2016)}]{Suhonen2016}%
  \BibitemOpen
  \bibfield  {author} {\bibinfo {author} {\bibfnamefont {P.~M.}\ \bibnamefont
  {Suhonen}}\ and\ \bibinfo {author} {\bibfnamefont {R.~P.}\ \bibnamefont
  {Linna}},\ }\href {https://doi.org/10.1103/physreve.93.012406} {\bibfield
  {journal} {\bibinfo  {journal} {Physical Review E}\ }\textbf {\bibinfo
  {volume} {93}} (\bibinfo {year} {2016})}\BibitemShut {NoStop}%
\bibitem [{\citenamefont {Yu}\ and\ \citenamefont {Luo}(2011)}]{Yu2011}%
  \BibitemOpen
  \bibfield  {author} {\bibinfo {author} {\bibfnamefont {W.}~\bibnamefont
  {Yu}}\ and\ \bibinfo {author} {\bibfnamefont {K.}~\bibnamefont {Luo}},\
  }\href {https://doi.org/10.1021/ja204892z} {\bibfield  {journal} {\bibinfo
  {journal} {Journal of the American Chemical Society}\ }\textbf {\bibinfo
  {volume} {133}},\ \bibinfo {pages} {13565} (\bibinfo {year}
  {2011})}\BibitemShut {NoStop}%
\bibitem [{\citenamefont {Muthukumar}(2001)}]{Muthukumar2001}%
  \BibitemOpen
  \bibfield  {author} {\bibinfo {author} {\bibfnamefont {M.}~\bibnamefont
  {Muthukumar}},\ }\href {https://doi.org/10.1103/physrevlett.86.3188}
  {\bibfield  {journal} {\bibinfo  {journal} {Physical Review Letters}\
  }\textbf {\bibinfo {volume} {86}},\ \bibinfo {pages} {3188} (\bibinfo {year}
  {2001})}\BibitemShut {NoStop}%
\bibitem [{\citenamefont {Cacciuto}\ and\ \citenamefont
  {Luijten}(2006)}]{Cacciuto2006}%
  \BibitemOpen
  \bibfield  {author} {\bibinfo {author} {\bibfnamefont {A.}~\bibnamefont
  {Cacciuto}}\ and\ \bibinfo {author} {\bibfnamefont {E.}~\bibnamefont
  {Luijten}},\ }\href {https://doi.org/10.1103/physrevlett.96.238104}
  {\bibfield  {journal} {\bibinfo  {journal} {Physical Review Letters}\
  }\textbf {\bibinfo {volume} {96}} (\bibinfo {year} {2006})}\BibitemShut
  {NoStop}%
\bibitem [{\citenamefont {Wong}\ and\ \citenamefont
  {Muthukumar}(2008)}]{Wong2008}%
  \BibitemOpen
  \bibfield  {author} {\bibinfo {author} {\bibfnamefont {C.~T.~A.}\
  \bibnamefont {Wong}}\ and\ \bibinfo {author} {\bibfnamefont {M.}~\bibnamefont
  {Muthukumar}},\ }\href {https://doi.org/10.1529/biophysj.108.135525}
  {\bibfield  {journal} {\bibinfo  {journal} {Biophysical Journal}\ }\textbf
  {\bibinfo {volume} {95}},\ \bibinfo {pages} {3619} (\bibinfo {year}
  {2008})}\BibitemShut {NoStop}%
\bibitem [{\citenamefont {Khalilian}\ \emph {et~al.}(2021)\citenamefont
  {Khalilian}, \citenamefont {Sarabadani},\ and\ \citenamefont
  {Ala-Nissila}}]{Khalilian2021}%
  \BibitemOpen
  \bibfield  {author} {\bibinfo {author} {\bibfnamefont {H.}~\bibnamefont
  {Khalilian}}, \bibinfo {author} {\bibfnamefont {J.}~\bibnamefont
  {Sarabadani}}, \ and\ \bibinfo {author} {\bibfnamefont {T.}~\bibnamefont
  {Ala-Nissila}},\ }\href {https://doi.org/10.1103/physrevresearch.3.013080}
  {\bibfield  {journal} {\bibinfo  {journal} {Physical Review Research}\
  }\textbf {\bibinfo {volume} {3}} (\bibinfo {year} {2021})}\BibitemShut
  {NoStop}%
\bibitem [{\citenamefont {Tan}\ \emph {et~al.}(2021)\citenamefont {Tan},
  \citenamefont {Chen},\ and\ \citenamefont {Zhao}}]{Tan2021}%
  \BibitemOpen
  \bibfield  {author} {\bibinfo {author} {\bibfnamefont {F.}~\bibnamefont
  {Tan}}, \bibinfo {author} {\bibfnamefont {Y.}~\bibnamefont {Chen}}, \ and\
  \bibinfo {author} {\bibfnamefont {N.}~\bibnamefont {Zhao}},\ }\href
  {https://doi.org/10.1039/d0sm01906b} {\bibfield  {journal} {\bibinfo
  {journal} {Soft Matter}\ }\textbf {\bibinfo {volume} {17}},\ \bibinfo {pages}
  {1940} (\bibinfo {year} {2021})}\BibitemShut {NoStop}%
\bibitem [{\citenamefont {jin Jeon}\ and\ \citenamefont
  {Muthukumar}(2016)}]{Jeon2016}%
  \BibitemOpen
  \bibfield  {author} {\bibinfo {author} {\bibfnamefont {B.}~\bibnamefont {jin
  Jeon}}\ and\ \bibinfo {author} {\bibfnamefont {M.}~\bibnamefont
  {Muthukumar}},\ }\href {https://doi.org/10.1021/acs.macromol.6b01663}
  {\bibfield  {journal} {\bibinfo  {journal} {Macromolecules}\ }\textbf
  {\bibinfo {volume} {49}},\ \bibinfo {pages} {9132} (\bibinfo {year}
  {2016})}\BibitemShut {NoStop}%
\bibitem [{\citenamefont {Nikoofard}\ and\ \citenamefont
  {Fazli}(2012)}]{Nikoofard2012}%
  \BibitemOpen
  \bibfield  {author} {\bibinfo {author} {\bibfnamefont {N.}~\bibnamefont
  {Nikoofard}}\ and\ \bibinfo {author} {\bibfnamefont {H.}~\bibnamefont
  {Fazli}},\ }\href {https://doi.org/10.1103/physreve.85.021804} {\bibfield
  {journal} {\bibinfo  {journal} {Physical Review E}\ }\textbf {\bibinfo
  {volume} {85}} (\bibinfo {year} {2012})}\BibitemShut {NoStop}%
\bibitem [{\citenamefont {Sharma}\ \emph {et~al.}(2022)\citenamefont {Sharma},
  \citenamefont {Kapri},\ and\ \citenamefont {Chaudhuri}}]{Sharma2022}%
  \BibitemOpen
  \bibfield  {author} {\bibinfo {author} {\bibfnamefont {A.}~\bibnamefont
  {Sharma}}, \bibinfo {author} {\bibfnamefont {R.}~\bibnamefont {Kapri}}, \
  and\ \bibinfo {author} {\bibfnamefont {A.}~\bibnamefont {Chaudhuri}},\ }\href
  {https://doi.org/10.1038/s41598-022-21845-6} {\bibfield  {journal} {\bibinfo
  {journal} {Scientific Reports}\ }\textbf {\bibinfo {volume} {12}} (\bibinfo
  {year} {2022})}\BibitemShut {NoStop}%
\bibitem [{\citenamefont {Lan}\ \emph {et~al.}(2011)\citenamefont {Lan},
  \citenamefont {Holden}, \citenamefont {Zhang},\ and\ \citenamefont
  {White}}]{Lan2011}%
  \BibitemOpen
  \bibfield  {author} {\bibinfo {author} {\bibfnamefont {W.-J.}\ \bibnamefont
  {Lan}}, \bibinfo {author} {\bibfnamefont {D.~A.}\ \bibnamefont {Holden}},
  \bibinfo {author} {\bibfnamefont {B.}~\bibnamefont {Zhang}}, \ and\ \bibinfo
  {author} {\bibfnamefont {H.~S.}\ \bibnamefont {White}},\ }\href
  {https://doi.org/10.1021/ac200312n} {\bibfield  {journal} {\bibinfo
  {journal} {Analytical Chemistry}\ }\textbf {\bibinfo {volume} {83}},\
  \bibinfo {pages} {3840} (\bibinfo {year} {2011})}\BibitemShut {NoStop}%
\bibitem [{\citenamefont {Sexton}\ \emph {et~al.}(2007)\citenamefont {Sexton},
  \citenamefont {Horne},\ and\ \citenamefont {Martin}}]{Sexton2007}%
  \BibitemOpen
  \bibfield  {author} {\bibinfo {author} {\bibfnamefont {L.~T.}\ \bibnamefont
  {Sexton}}, \bibinfo {author} {\bibfnamefont {L.~P.}\ \bibnamefont {Horne}}, \
  and\ \bibinfo {author} {\bibfnamefont {C.~R.}\ \bibnamefont {Martin}},\
  }\href {https://doi.org/10.1039/b708725j} {\bibfield  {journal} {\bibinfo
  {journal} {Molecular {BioSystems}}\ }\textbf {\bibinfo {volume} {3}},\
  \bibinfo {pages} {667} (\bibinfo {year} {2007})}\BibitemShut {NoStop}%
\bibitem [{\citenamefont {Siwy}\ and\ \citenamefont
  {Fuli{\'{n}}ski}(2002)}]{Siwy2002}%
  \BibitemOpen
  \bibfield  {author} {\bibinfo {author} {\bibfnamefont {Z.}~\bibnamefont
  {Siwy}}\ and\ \bibinfo {author} {\bibfnamefont {A.}~\bibnamefont
  {Fuli{\'{n}}ski}},\ }\href {https://doi.org/10.1103/physrevlett.89.198103}
  {\bibfield  {journal} {\bibinfo  {journal} {Physical Review Letters}\
  }\textbf {\bibinfo {volume} {89}} (\bibinfo {year} {2002})}\BibitemShut
  {NoStop}%
\bibitem [{\citenamefont {Sung}\ and\ \citenamefont {Park}(1996)}]{Sung1996}%
  \BibitemOpen
  \bibfield  {author} {\bibinfo {author} {\bibfnamefont {W.}~\bibnamefont
  {Sung}}\ and\ \bibinfo {author} {\bibfnamefont {P.~J.}\ \bibnamefont
  {Park}},\ }\href {https://doi.org/10.1103/physrevlett.77.783} {\bibfield
  {journal} {\bibinfo  {journal} {Physical Review Letters}\ }\textbf {\bibinfo
  {volume} {77}},\ \bibinfo {pages} {783} (\bibinfo {year} {1996})}\BibitemShut
  {NoStop}%
\bibitem [{\citenamefont {Muthukumar}(1999)}]{Muthukumar1999}%
  \BibitemOpen
  \bibfield  {author} {\bibinfo {author} {\bibfnamefont {M.}~\bibnamefont
  {Muthukumar}},\ }\href {https://doi.org/10.1063/1.480386} {\bibfield
  {journal} {\bibinfo  {journal} {The Journal of Chemical Physics}\ }\textbf
  {\bibinfo {volume} {111}},\ \bibinfo {pages} {10371} (\bibinfo {year}
  {1999})}\BibitemShut {NoStop}%
\bibitem [{\citenamefont {Muthukumar}(2003)}]{Muthukumar2003}%
  \BibitemOpen
  \bibfield  {author} {\bibinfo {author} {\bibfnamefont {M.}~\bibnamefont
  {Muthukumar}},\ }\href {https://doi.org/10.1063/1.1553753} {\bibfield
  {journal} {\bibinfo  {journal} {The Journal of Chemical Physics}\ }\textbf
  {\bibinfo {volume} {118}},\ \bibinfo {pages} {5174} (\bibinfo {year}
  {2003})}\BibitemShut {NoStop}%
\bibitem [{\citenamefont {Guttmann}\ and\ \citenamefont
  {Torrie}(1984)}]{Guttmann1984}%
  \BibitemOpen
  \bibfield  {author} {\bibinfo {author} {\bibfnamefont {A.~J.}\ \bibnamefont
  {Guttmann}}\ and\ \bibinfo {author} {\bibfnamefont {G.~M.}\ \bibnamefont
  {Torrie}},\ }\href {https://doi.org/10.1088/0305-4470/17/18/023} {\bibfield
  {journal} {\bibinfo  {journal} {Journal of Physics A: Mathematical and
  General}\ }\textbf {\bibinfo {volume} {17}},\ \bibinfo {pages} {3539}
  (\bibinfo {year} {1984})}\BibitemShut {NoStop}%
\bibitem [{\citenamefont {Kumar}\ \emph {et~al.}(2020)\citenamefont {Kumar},
  \citenamefont {Chauhan}, \citenamefont {Singh},\ and\ \citenamefont
  {Foster}}]{Kumar2020}%
  \BibitemOpen
  \bibfield  {author} {\bibinfo {author} {\bibfnamefont {S.}~\bibnamefont
  {Kumar}}, \bibinfo {author} {\bibfnamefont {K.}~\bibnamefont {Chauhan}},
  \bibinfo {author} {\bibfnamefont {S.}~\bibnamefont {Singh}}, \ and\ \bibinfo
  {author} {\bibfnamefont {D.}~\bibnamefont {Foster}},\ }\href
  {https://doi.org/10.1103/physreve.101.030502} {\bibfield  {journal} {\bibinfo
   {journal} {Physical Review E}\ }\textbf {\bibinfo {volume} {101}} (\bibinfo
  {year} {2020})}\BibitemShut {NoStop}%
\bibitem [{\citenamefont {Chauhan}\ and\ \citenamefont
  {Kumar}(2021)}]{Chauhan2021}%
  \BibitemOpen
  \bibfield  {author} {\bibinfo {author} {\bibfnamefont {K.}~\bibnamefont
  {Chauhan}}\ and\ \bibinfo {author} {\bibfnamefont {S.}~\bibnamefont
  {Kumar}},\ }\href {https://doi.org/10.1103/physreve.103.042501} {\bibfield
  {journal} {\bibinfo  {journal} {Physical Review E}\ }\textbf {\bibinfo
  {volume} {103}} (\bibinfo {year} {2021})}\BibitemShut {NoStop}%
\bibitem [{\citenamefont {Vanderzande}(1998)}]{Vanderzande1998}%
  \BibitemOpen
  \bibfield  {author} {\bibinfo {author} {\bibfnamefont {C.}~\bibnamefont
  {Vanderzande}},\ }\href {https://doi.org/10.1017/cbo9780511563935} {\emph
  {\bibinfo {title} {Lattice Models of Polymers}}}\ (\bibinfo  {publisher}
  {Cambridge University Press},\ \bibinfo {year} {1998})\BibitemShut {NoStop}%
\bibitem [{\citenamefont {Kumar}\ and\ \citenamefont {Li}(2010)}]{Kumar2010}%
  \BibitemOpen
  \bibfield  {author} {\bibinfo {author} {\bibfnamefont {S.}~\bibnamefont
  {Kumar}}\ and\ \bibinfo {author} {\bibfnamefont {M.~S.}\ \bibnamefont {Li}},\
  }\href {https://doi.org/10.1016/j.physrep.2009.11.001} {\bibfield  {journal}
  {\bibinfo  {journal} {Physics Reports}\ }\textbf {\bibinfo {volume} {486}},\
  \bibinfo {pages} {1} (\bibinfo {year} {2010})}\BibitemShut {NoStop}%
\bibitem [{\citenamefont {Sun}\ \emph {et~al.}(2009)\citenamefont {Sun},
  \citenamefont {Cao},\ and\ \citenamefont {Luo}}]{Sun2009}%
  \BibitemOpen
  \bibfield  {author} {\bibinfo {author} {\bibfnamefont {L.-Z.}\ \bibnamefont
  {Sun}}, \bibinfo {author} {\bibfnamefont {W.-P.}\ \bibnamefont {Cao}}, \ and\
  \bibinfo {author} {\bibfnamefont {M.-B.}\ \bibnamefont {Luo}},\ }\href
  {https://doi.org/10.1063/1.3264944} {\bibfield  {journal} {\bibinfo
  {journal} {The Journal of Chemical Physics}\ }\textbf {\bibinfo {volume}
  {131}},\ \bibinfo {pages} {194904} (\bibinfo {year} {2009})}\BibitemShut
  {NoStop}%
\end{thebibliography}%

\end{document}